\documentclass[runningheads]{llncs}
\usepackage[T1]{fontenc}
\usepackage{times}
\usepackage{epsfig}
\usepackage{graphicx}
\usepackage{amsmath}
\usepackage{amssymb}
\usepackage{natbib}
\usepackage{latexsym}
\usepackage{microtype}
\usepackage{inconsolata}
\usepackage{latexsym}
\usepackage{xcolor}
\usepackage{makecell}
\usepackage{latexsym}
\usepackage{hyperref}
\usepackage{inconsolata}
\usepackage{url}
\usepackage{makecell}
\usepackage{multirow}
\usepackage{graphicx}
\usepackage{caption}
\usepackage{subcaption}
\usepackage{amsmath}
\usepackage{bbm}
\usepackage{booktabs}
\usepackage{algorithmic}
\usepackage{amssymb}
\usepackage{bm}
\usepackage[ruled,vlined,linesnumbered]{algorithm2e}
\usepackage{enumitem}
\usepackage{pifont}
\usepackage{makecell}
\usepackage{color}

\usepackage{indentfirst}
\usepackage{graphicx}
\begin{document}
\title{DeepFM-Crispr: Prediction of CRISPR On-Target Effects via Deep Learning
}
%
%
\author{Condy Bao$^1$\thanks{Corresponding author: \url{condybao@gmail.com}}~\,  Fuxiao Liu$^2$}
%
%
\institute{$^1$1St Mark’s School, USA, $^2$University of Maryland, USA}
%
\maketitle              
\begin{abstract}

Since the advent of CRISPR-Cas9, a groundbreaking gene-editing technology that enables precise genomic modifications via a short RNA guide sequence, there has been a marked increase in the accessibility and application of this technology across various fields. The success of CRISPR-Cas9 has spurred further investment and led to the discovery of additional CRISPR systems, including CRISPR-Cas13. Distinct from Cas9, which targets DNA, Cas13 targets RNA, offering unique advantages for gene modulation. We focus on Cas13d, a variant known for its collateral activity where it non-specifically cleaves adjacent RNA molecules upon activation, a feature critical to its function. We introduce DeepFM-Crispr, a novel deep learning model developed to predict the on-target efficiency and evaluate the off-target effects of Cas13d. This model harnesses a large language model to generate comprehensive representations rich in evolutionary and structural data, thereby enhancing predictions of RNA secondary structures and overall sgRNA efficacy. A transformer-based architecture processes these inputs to produce a predictive efficacy score. Comparative experiments show that DeepFM-Crispr not only surpasses traditional models but also outperforms recent state-of-the-art deep learning methods in terms of prediction accuracy and reliability.

\end{abstract}
\section{Introduction}
The discovery and development of Clustered Regularly Interspaced Short Palindromic Repeats (CRISPR) and their associated Cas proteins have revolutionized biotechnology and biomedical sciences. Initially identified within the adaptive immune systems of bacteria and archaea, these CRISPR-Cas systems have been ingeniously adapted for genome editing. They exploit their inherent ability to make precise and efficient genetic alterations \cite{doudna346genome, jinek2012programmable}. The CRISPR locus is characterized by repetitive base sequences interspersed with spacers derived from past viral and plasmid invaders. This locus is transcribed into a long precursor CRISPR RNA (pre-crRNA), which is then processed into mature guide RNAs (crRNAs). These crRNAs direct the Cas proteins to cleave complementary sequences in invading genetic elements, thereby providing adaptive immunity. Among the various CRISPR-Cas systems, CRISPR-Cas9 of Type II has attracted significant attention due to its simplicity and versatility in genome editing. It employs a dual-RNA structure consisting of crRNA and trans-activating crRNA (tracrRNA), guiding the Cas9 enzyme to specific DNA targets \cite{jinek2012programmable, doudna2014new}.



CRISPR-Cas12, a Type V system, offers unique advantages over Cas9 by utilizing a single RNA for both CRISPR array processing and target DNA recognition. This system cleaves target DNA in a staggered manner, increasing the diversity of editable sequences and enhancing the potential for multiplex editing. Another innovative addition to the CRISPR toolkit is CRISPR-Cas13, which targets RNA instead of DNA, facilitating not only gene modulation without altering the genome but also enabling novel diagnostic applications due to its collateral cleavage activity upon target recognition \cite{gootenberg2017nucleic}.  The specificity and efficacy of guide RNA design are crucial for maximizing on-target actions and minimizing off-target effects, which are particularly concerning with Cas9’s potential DNA off-targets and Cas13’s RNA-targeted collateral activity \cite{abudayyeh2017rna}.  CRISPR screens, including those for viability and FACS-sorting, have become instrumental in evaluating the effectiveness and specificity of CRISPR systems. These screens employ a multitude of guide RNAs to ascertain factors influencing knockout efficiency and to refine guide designs to balance activity and specificity \cite{doench2016optimized}.  Unlike tools focused solely on CRISPR-Cas9, those designed for Cas13 must account for RNA secondary structures, significantly impacting guide RNA efficiency \cite{wessels2020massively}.


To address challenges related to data scarcity and the complexity of integrating structural and evolutionary information, we introduce DeepFM-Crispr. This model leverages advanced transformer-based architectures and large language models, which have revolutionized fields ranging from natural language processing \cite{liu2023aligning,liu2024large,liu2023mmc,li2023towards} to bioinformatics due to their ability to handle large datasets and extract deep, contextual relationships within data. Our transformer \cite{vaswani2017attention} model processes a wide array of inputs, offering a predictive score for on-target efficiency based on enriched representations of evolutionary and structural insights. Validated against a dataset of 22,599 Cas13d sgRNAs \cite{cheng2023modeling}, DeepFM-Cas13d outperforms conventional machine learning methods and existing prediction tools, particularly in targeting non-coding RNAs \cite{shmakov2015discovery}.

\begin{figure*}[t]
    \centering
      \includegraphics[width=1\textwidth]{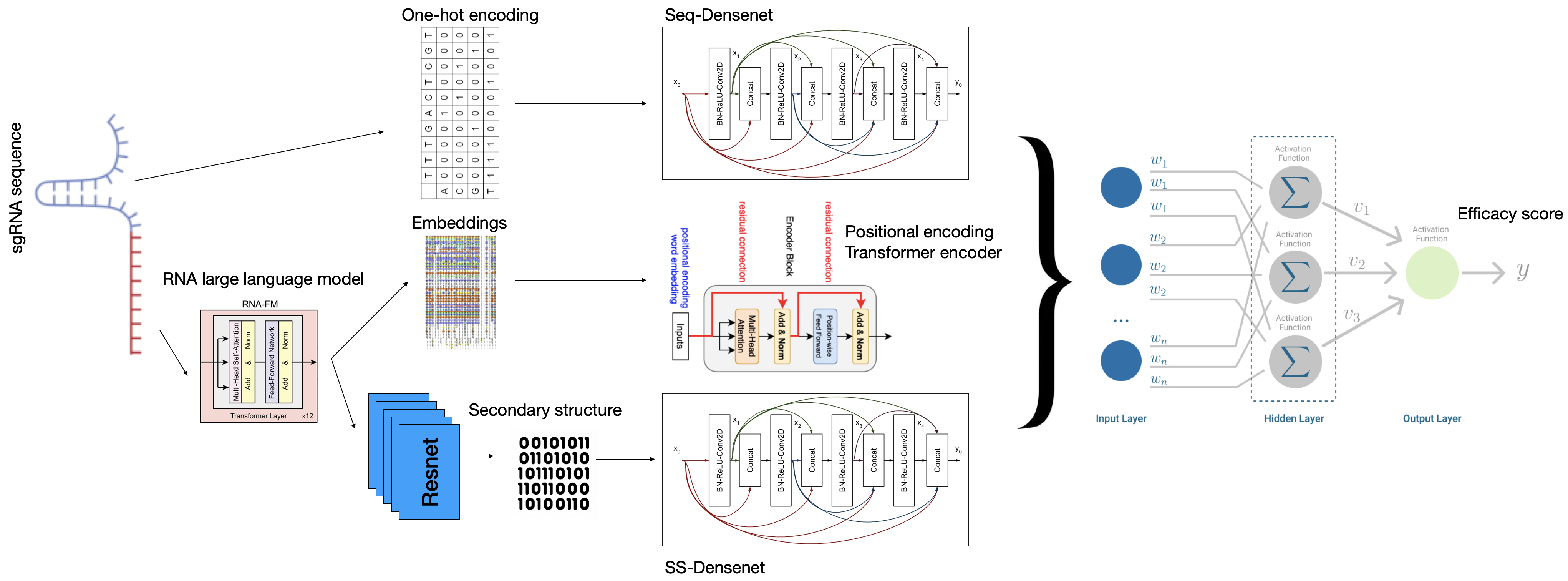}
    \caption{\textbf{The Overall Architecture of \textit{DeepFM-Crispr}.}}
    \vspace{-0.1in}
    \label{fig:model}
\end{figure*}

\section{Method}
Our methodology integrates multiple deep learning architectures and data representation techniques to predict the on-target efficiency of sgRNAs based on their sequences and secondary structures. This approach harnesses the power of large language models, convolutional networks, and transformer encoders to process and analyze the complex biological data. The details of DeepFM-Crispr shown in Fig. \ref{fig:model}.

\subsection{Data Representation}
sgRNA sequences were encoded using a one-hot encoding scheme, where each nucleotide (A, C, G, U) is represented by a binary vector. The vectors for adenine (A), cyto- sine (C), guanine (G), and uracil (U) are respectively [1,0,0,0], [0,1,0,0], [0,0,1,0], and [0,0,0,1].  This methodological choice ensures a uniform input structure for all sequences, facilitating the computational handling of genetic data across diverse sgRNA samples.  By transforming the nucleotide sequences into binary vectors, the model can effectively learn from the positional and compositional nuances of the sgRNA without the biases and variances inherent in raw textual data.

The one-hot encoded vectors serve as the primary input for the subsequent layers of the machine learning architecture. They are fed into a series of deep learning models that are designed to extract and learn complex patterns and relationships.  This initial representation forms the basis for all further transformations and feature extractions performed by the RNA large language model and other components of our predictive framework.  The standardized format ensures that each sgRNA is represented in a consistent manner, allowing the deep learning algorithms to focus on learning the underlying biological mechanisms rather than adjusting to variations in data format.

\subsection{RNA Large Language Model}
RNA-FM \cite{chen2022interpretable} is designed as an end-to-end deep learning model that efficiently extracts latent features from RNA sequences and leverages an attention mechanism to capture contextual information.  It features 12 layers of transformer-based bidirectional encoder blocks equipped with positional embeddings.  This structure allows RNA-FM to accurately discern the positional context of ncRNA sequences.  The encoder within RNA-FM utilizes self-attention and feedforward connections to generate complex representations that integrate context from every sequence position. Furthermore, the model is adept at constructing pairwise interactions between nucleotides, enhancing its ability to depict direct nucleotide-nucleotide interactions and providing a nuanced representation of the input data.  These capabilities make RNA-FM particularly effective in correlating internal representations with RNA secondary structures.  As a result, the model produces high-dimensional embeddings for each sgRNA \cite{abudayyeh2017rna}, encapsulating both local and global contextual relationships within the sequences.  These embeddings are subsequently utilized as inputs for further predictive modeling to determine sgRNA efficacy scores, and also serve as inputs for a secondary structure prediction model, enhancing our understanding and prediction of RNA structural configurations.

\subsection{Secondary Structure Prediction}
The secondary structure of each sgRNA was predicted using a ResNet model \cite{krishna2017dense} that processes inputs derived from the RNA-FM representations.   This  model outputs  a probability matrix where each nucleotide position is labeled  as either paired  (1)  or unpaired (0). These binary sequences are then further processed using a deep convolutional network based on the ResNet  architecture,  which is particularly adept at capturing spatial hierarchies in structured data \cite{he2016deep}.  This approach allows for effective integration of contextual and spatial information, enhancing the accuracy of the secondary structure predictions for the sgRNAs.

\subsection{Feature Integration and Processing}
The embeddings  from the  RNA-FM and the outputs from the secondary structure prediction ResNet are integrated and further processed to refine the feature representation.

\textbf{DenseNet architecture.} Seq-DenseNet \cite{wei2021self} and SS-DenseNet \cite{zhao2021hyperspectral}, was employed to process integrated features. This architecture benefits from dense connectivity patterns that improve the flow of information and gradients throughout the network, aiding in the robust learning of features from both sequence and structural data \cite{wei2021self}. 

\textbf{Positional Encoding Transformer Encoder.} The output embedding of RNA-FM are passed through a positional encoding transformer encoder. This module incorporates positional encodings to the input features to maintain the sequence order, which is crucial for capturing dependencies that are positionally distant in the sgRNA sequence. The transformer encoder refines these features by focusing on the most relevant parts of the sgRNA for efficacy prediction \cite{huang2017densely}.

\subsection{Efficacy Prediction}
The final prediction of sgRNA efficacy is performed using a multi-layer perceptron (MLP) \cite{riedmiller2014multi}. The MLP comprises an input layer that receives the processed features, several hidden layers with non-linear activation functions to capture complex relationships in the data, and an output layer that produces a continuous efficacy score for each sgRNA.

\begin{figure*}[t]
    \centering
      \includegraphics[width=1\textwidth]{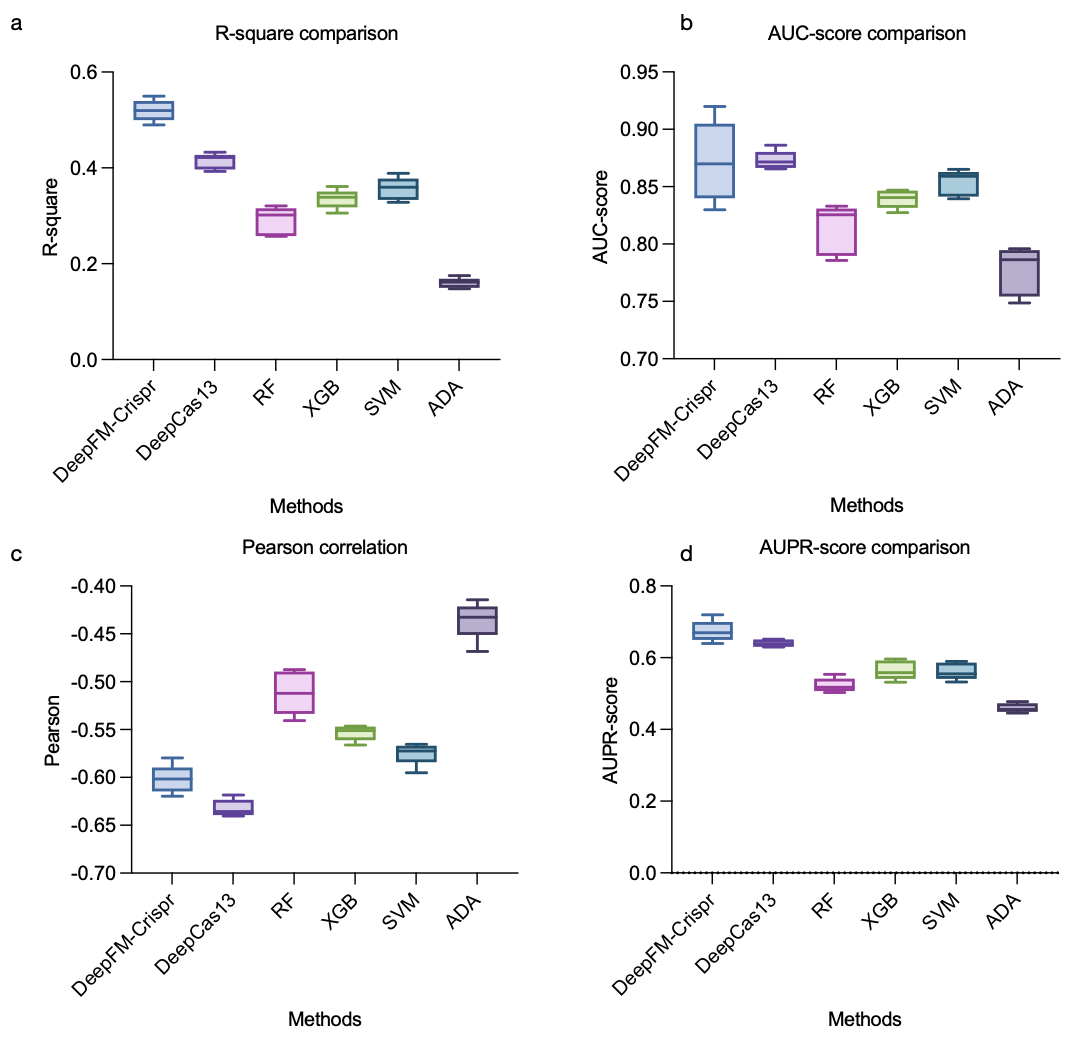}
    \caption{\textbf{Comparison Result between \textit{DeepFM-Crispr} with baselines. }}
    \vspace{-0.1in}
    \label{fig:result}
\end{figure*}

\section{Dataset}

To explore the efficiency and specificity of the Cas13d system, we performed a comprehensive two-vector CRISPR/Cas13d proliferation screen on the A375 melanoma cell line, guided by protocols from a well-established benchmark study \cite{cheng2023modeling}. The screening library consisted of 10,830 sgRNAs targeting a total of 426 genes, including 192 protein-coding genes and 234 long non-coding RNAs (lncRNAs). This selection included 94 essential genes and 14 non-essential genes, previously identified in A375 cells via RNA interference and CRISPR-based screenings. Our library design aimed to robustly model the efficiencies of Cas13d sgRNAs, allocating approximately 30 guides per gene. This strategy was intended to overcome potential biases observed in previous tiling screens that were limited to only 2-3 genes, thereby enhancing the representativity of the sgRNA library.

Following the screening, sgRNA abundance was quantitatively assessed through high-throughput sequencing. Data analysis was conducted using the MAGeCK algorithm to evaluate sgRNA performance and gene essentiality. Quality control checks were stringent, with an average of 5.6 million reads per sample, ensuring reliable data capture. Each guide was represented by over 300 reads, with fewer than four guides missing per gene—indicative of comprehensive coverage. The Gini coefficient was maintained below 0.06, confirming a uniform and non-biased distribution of guide representation across the dataset. Most notably, 20 of the essential genes demonstrated significant depletion, with a false discovery rate (FDR) of less than 10\%, underscoring the screening's effectiveness in identifying gene functionalities crucial for cell proliferation in melanoma.

\section{Experiments}

\subsection{Implementation}

\textbf{Batch Size:} The model was trained with a batch size of 128 sgRNAs to balance computational efficiency with memory constraints.

\textbf{Learning Rate:} We employed a learning rate of 1e-4, utilizing a learning rate scheduler to decrease the rate by 10\% every two epochs to stabilize training as it progressed.

\textbf{Epochs:} The model was trained for up to 50 epochs, with early stopping implemented if the validation loss did not improve for 5 consecutive epochs. This approach prevented overfitting and ensured that the model generalized well to new, unseen data.

\textbf{Optimizer:} The Adam optimizer \cite{zhang2018improved} was used for its adaptive learning rate capabilities, which helped in converging faster and more effectively than traditional stochastic gradient descent. L2 regularization \cite{cortes2012l2} was applied to all trainable parameters to prevent overfitting by penalizing large weights. A dropout rate of 0.1 was used in each transformer layer to randomly omit a subset of features during training, further helping the model to generalize better. To handle the varying lengths of sgRNA sequences, attention masks were used within the transformer layers to ignore padding tokens during the self-attention calculations.

\textbf{Baseline Methods:} In the evaluation of guide RNA (gRNA) efficacy prediction, the DeepFM-Crispr model was compared against several established machine learning methods, including Random Forest (RF) \cite{rigatti2017random}, XGBoost (XGB) \cite{chen2016xgboost}, Support Vector Machine (SVM) \cite{suthaharan2016support}, AdaBoost (ADA) \cite{ying2013advance}, and a recent deep learning method,  DeepCas13 \cite{wessels2020massively}.  

\subsection{Evaluation}
Conventional  machine  learning  algorithms  were applied using 185 curated features consistent with previous studies.  Each model was trained and validated on three publicly available Cas13d tiling screening datasets, encompassing a total of 5,726 sgRNAs, employing five-fold cross-validation to ensure robustness. The evaluation of our models focused on two primary metrics: (1) the prediction accuracy of sgRNA efficacy across the entire dataset, and (2) the ability to classify sgRNAs into efficient or non-efficient categories based on their performance.

\subsection{Result Discussion}

\textbf{Prediction Accuracy.} The first metric of evaluation was the coefficient of determination (R²) and the Pearson correlation coefficient (PCC) between the predicted efficacy scores and the actual log fold changes (LFCs). DeepFM-Crispr demonstrated superior performance in this regard, achieving a higher R² value and a more pronounced negative Pearson correlation. These results, illustrated in Fig. \ref{fig:result}, indicate that DeepFM-Crispr provides more accurate predictions of sgRNA efficacy, aligning closely with experimental outcomes.

\textbf{Classification of sgRNA Efficiency.} For the classification task, sgRNAs were categorized based on their LFC values; those with LFC smaller than -0.5 were classified as positive (efficient), and all others as negative (non-efficient). The effectiveness of each model in this binary classification was measured by the area under the Receiver Operating Characteristic (ROC) curve (AUC) and the area under the precision-recall curve (AUPR). DeepFM-Crispr not only matched the top AUC performance of DeepCas13 at an average of 0.88 across five-fold cross-validation (as shown in Fig. \ref{fig:result}) but also significantly outperformed other methods, which exhibited AUC scores ranging from 0.78 to 0.85. Furthermore, DeepFM-Crispr excelled in the precision-recall metric, achieving an average AUPR score of 0.69. This score was notably higher than those achieved by DeepCas13 and other traditional approaches, which varied between 0.45 and 0.58 (depicted in Fig. \ref{fig:result}). This indicates a stronger capability of DeepFM-Crispr to differentiate between strong and weak knockdown effects. The high AUPR score is particularly significant in the context of the dataset’s imbalance, where positive samples are less frequent, underscoring the model’s robustness in distinguishing positives from negatives.

\textbf{Implications for Gene Editing Applications.} These findings underscore the enhanced predictive accuracy of DeepFM-Crispr in assessing sgRNA efficacy, affirming its utility in gene editing applications where precise guide RNA selection is crucial. The ability of DeepFM-Crispr to accurately predict and classify sgRNA efficiency supports its potential as a valuable tool in optimizing CRISPR-based gene editing. This is particularly in therapeutic contexts where the precision of genetic modification can dictate treatment efficacy.

\section{Related Work}
The CRISPR-Cas systems \cite{doudna346genome, jinek2012programmable} have revolutionized the field of genetic engineering, offering unprecedented precision in gene editing \cite{jinek2012programmable, doudna2014new, khalil2020genome}. Since its inception, the CRISPR-Cas9 system has been extensively studied and applied across various biological contexts due to its ability to make targeted DNA modifications. However, the discovery of CRISPR-Cas13 \cite{abudayyeh2017rna}, which targets RNA, has opened new avenues for gene modulation without altering the DNA itself. Among the variants, Cas13d is particularly notable for its collateral activity, where it cleaves nearby non-target RNA sequences upon activation, offering potential for diagnostic as well as therapeutic applications \cite{barrangou2007crispr}.

Recent advancements have leveraged computational models to enhance the predictability and efficiency of CRISPR systems. For instance, models like CRISPRpred-seq \cite{muhammad2020crisprpred} and DeepCas \cite{wessels2020massively} have utilized traditional machine learning and deep learning techniques \cite{li2024mosaic,fei2024multimodal,wang2024mementos} to predict sgRNA efficacy, primarily focusing on CRISPR-Cas9. These models often rely on sequence-based features and have shown significant promise in reducing off-target effects and enhancing on-target efficiency. However, the unique mechanisms and RNA targeting properties of Cas13d \cite{gupta2022cas13d} present distinct challenges and opportunities that these models are not tailored to address.

DeepFM-Crispr introduces an innovative approach by integrating large language models and transformer-based architectures to specifically enhance the performance of CRISPR-Cas13d systems. This model surpasses traditional and recent computational approaches by effectively capturing and processing extensive evolutionary and structural information pertinent to RNA. The use of a large language model allows DeepFM-Crispr to understand and predict the complex dynamics of RNA interactions \cite{cheng2023modeling}, which are crucial for achieving high precision in RNA-targeted gene editing. Our model not only builds upon the existing body of knowledge but also sets a new benchmark for computational tools in the CRISPR field, particularly for RNA-targeting systems. By focusing on Cas13d, DeepFM-Crispr addresses a critical gap in the existing tools, offering refined predictions and insights that are vital for both research and therapeutic applications.

\section{Conclusion}

The DeepFM-Crispr model has proven to be exceptionally versatile and robust, demonstrating not only its efficacy with the Cas13d system but also its potential applicability to other CRISPR-Cas systems. Leveraging advanced large language model techniques, DeepFM-Crispr adeptly captures complex genetic interactions and sequence nuances essential for precise genome editing. The model's strength lies in its ability to integrate sophisticated deep learning techniques with large-scale genomic data, enabling it to surpass traditional models in both predictive accuracy and operational efficiency. This adaptability positions DeepFM-Crispr as a valuable asset across a broad spectrum of CRISPR technologies, potentially revolutionizing gene editing methodologies across diverse applications.

\section{Future Work}
Looking forward, our research will extend the application of DeepFM-Crispr to other widely used CRISPR systems \cite{barrangou2016applications,wu2024safety}, such as Cas9 and Cas12. These systems play crucial roles in both basic research and clinical settings. By tailoring DeepFM-Crispr to these platforms, we aim to enhance the specificity and efficiency of sgRNA design, thereby minimizing off-target effects and optimizing therapeutic outcomes. Future developments will also focus on refining the model’s architecture to incorporate a more comprehensive range of genomic features and environmental factors, which are expected to further enhance its predictive capabilities.

As CRISPR technology continues to evolve, the integration of sophisticated computational models like DeepFM-Crispr is poised to drive significant advancements in the field of genetic engineering. This synergy between cutting-edge biotechnology and computational innovation opens new avenues for medical research and treatment strategies, holding promise for transformative impacts on healthcare and disease management.

\bibliographystyle{ACM-Reference-Format}
\bibliography{main}
\end{document}